**Single-cell protein dynamics reproduce universal fluctuations in cell populations**

Naama Brenner,*[†] Erez Braun,[‡ †] Anna Yoney, [§] Lee Susman,[$] James Rotella,[§] and Hanna Salman[§#]

**Department of Chemical Engineering, [†]Laboratory of Network Biology, [‡]Department of Physics, [$]Department of Mathematics, Technion, Haifa 32000, Israel*

*[§] Department of Physics and Astronomy, [#]Department of Computational and Systems Biology, University of Pittsburgh, Pittsburgh, PA 15260, USA*

Address reprint requests and inquiries to: nbrenner@tx.technion.ac.il, and hsalman@pitt.edu

ABSTRACT Protein variability in single cells has been studied extensively in populations, but little is known about temporal protein fluctuations in a single cell over extended times. We present here traces of protein copy number measured in individual bacteria over multiple generations and investigate their statistical properties, comparing them to previously measured population snapshots. We find that temporal fluctuations in individual traces exhibit the same universal features as those previously observed in populations. Scaled fluctuations around the mean of each trace exhibit the same universal distribution shape as found in populations measured under a wide range of conditions and in two distinct microorganisms. Additionally, the mean and variance of the traces over time obey the same quadratic relation. Analyzing the temporal features of the protein traces in individual cells, reveals that within a cell cycle protein content increases as an exponential function with a rate that varies from cycle to cycle. This leads to a compact description of the protein trace as a 3-variable stochastic process—the exponential rate, the cell-cycle duration and the value at the cycle start—sampled once each cell cycle. This compact description is sufficient to preserve the universal statistical properties of the protein fluctuations, namely, the protein distribution shape and the quadratic relationship between variance and mean. Our results show that the protein distribution shape is insensitive to sub-cycle intracellular microscopic details and reflects global cellular properties that fluctuate between generations.

## Introduction

The protein content of biological cells is a major determinant of their metabolism, growth and functionality. Despite its important role in shaping the phenotype, it is well established that the protein copy number varies widely among individuals in a cell population, even for highly expressed proteins in genetically identical cells grown under uniform conditions. Often interpreted as noise in gene expression, protein variation has attracted much attention, with the aim of understanding its biological significance and as a probe of the underlying molecular processes [1–4]. Utilizing the advancement of single-cell experimental techniques, in particular applied to microbial populations as model systems, protein variation was measured under a wide range of conditions [5–7]. Apart from special cases such as extremely low protein copy number or specific circuits giving rise to bimodality [8,9], the general characteristics emerging are that protein distributions are unimodal, broad, skewed and highly non-Gaussian [5,7,10–13].

Many intracellular processes have been identified as contributing to variation in protein copy-number. These include the plethora of molecular processes directly underlying protein production and its regulation, but also other more global cellular processes coupled to them such as metabolism and cell division. Indeed, much effort has been devoted to characterizing these various specific processes and their stochastic nature, including gene expression [6, 12, 15–21], cell division [18], growth rate [22] and more. Special emphasis has recently been placed on the contribution of promoter architecture to protein variation, with synthetic biology providing tools to isolate this contribution from other cellular processes [23,24]. The results of these studies reveal a range of different behaviors depending on context.

However, despite much advance in identifying and characterizing specific mechanisms that contribute to protein variation, their integration resulting in the total variation remains poorly understood. We have recently developed a phenomenological approach to investigate systematically the sensitivity of protein distributions to underlying biological processes [11]. We have probed the effect of an array of experimental control parameters, known to affect protein expression in cells (different proteins and different metabolic conditions), on the protein variation. Our results showed that when viewed in appropriately scaled variables (subtracting the mean and dividing by the standard deviation), all distributions from the entire array of experiments collapsed to the same non-Gaussian universal curve. The universal nature of these distributions suggests that they are not dominated by specific microscopic stochastic events.

Moreover, for all these measurements the variance scaled quadratically with the mean, implying that a single population-average measurement is enough to reconstruct the distribution in physical units. The range of control parameters leading to this universal behavior rendered this result significant but its source and limits of validity remained unclear.

It is important to remember that a population of dividing microorganisms is not an ensemble of independent particles, but a stochastic dynamical system far from equilibrium: proteins and other molecules are constantly being produced and degraded; at the same time, cells continuously grow and divide and their resources are passed along generations. The process of cell division is tightly coupled to cell growth and metabolism and incorporates both deterministic and stochastic components [25,26]. Following division, each cell starts its life-cycle with a phenotypic inheritance which provides the initial condition for its subsequent growth. The dynamic processes of protein production, cell division and inheritance are all crucial components in the building up of phenotypic variation in a population. Therefore it is of highest interest to measure these dynamics directly at the single-cell level over multiple generations.

Nonetheless, reviewing the large literature on protein variation one finds that practically all previous experiments were carried out on large cell populations measured at a given point in time. This provides a snapshot sample (some dynamical aspects can be probed by performing consecutive snapshots [27,28]), but direct measurements of protein content at single-cell resolution over extended timescales have not yet been carried out. In contrast to statistical physical systems at equilibrium, where ergodicity ensures that measurements over an ensemble are equivalent to measurements over times in an individual, in a biological population this is far from trivial and a range of behaviors may be expected. At one limit there may be ergodicity, with long-term measurements over a single isolated cell reproducing the same distributions as in a well-mixed cell population. At the other limit, collective effects and sensitivity to history and environment may dominate and lead to very different distributions in single cells over time versus population sampling.

In this study we present protein traces measured directly in single isolated bacterial cells, using a special experimental system designed for this purpose, and followed over extended times that cover multiple cycles of growth and division. The extended timescale of the experiment allows, for the first time, to collect a faithful sample of statistical properties over time in single cells; not only low moments but the full protein distribution can be characterized for each trace separately. These data enable us to investigate how the statistical properties of the population at a given time relate to the long-term single-cell fluctuations over their lifetime. We do this by comparing the statistical properties of protein fluctuations measured in our previous experiments on populations to those newly measured in individual bacteria over time. Our goal is to determine to what extent an individual cell samples over its lifetime the same distribution seen in a population. In particular, we are interested in examining whether the universal properties of protein fluctuations reported previously for a cell population, namely universal shape in scaled units and quadratic relationship between variance and mean [11], are observed in the temporal fluctuations in single cell as well. As will be shown below, we find that the relationship between temporal and population statistics is far from trivial but does contain information about the relevant timescales and processes that play a role in determining the fluctuations distribution.

**Results**

To access the temporal dimension of protein variation we have developed a microfluidic device to trap single *E. coli* cells and follow their size, division and protein content over extended times on the order of ~150 hours (~70 generations) (see Fig. 1 and Methods). A similar experimental system was used to study aging and cell division by following the dynamics of cell size [29]. Here, we concentrate on the variation in protein content which we can directly compare to population distributions [11]. In our experiments, cellular protein content was measured by the fluorescence intensity of a green fluorescent protein (GFP) regulated by three different promoters (see Methods for details). The environmental conditions (temperature and growth medium) are similar to our previous experiments on populations [11] and probe the protein content in the regime where its copy number is relatively high; genome-wide studies in both bacteria [7] and yeast [6] have shown that the majority of cellular proteins are in this regime. Moreover we are again interested in the integrated variation as contributed from multiple cellular processes, therefore we compare proteins that are metabolically relevant with ones that do not participate in growth metabolism. The universal distribution in populations was found to be insensitive to whether the protein is metabolically relevant or simply a marker [11]. Nevertheless, for the protein dynamics in single dividing cell over time this issue needs to be carefully measured. For example, for studying variation in one of the LAC operon proteins which are essential for lactose utilization, it is important that the cells are grown in a medium containing lactose as the main carbon source, thus ensuring that the expression process is metabolically relevant and coupled to all other cellular processes. For comparison a foreign viral promoter is also studied.

Typical measurements of cell length and fluorescence level reflecting protein content in a single trapped bacterium are shown in Figs. 1C and 1D, respectively (see also Fig. S1). Comparing the averages of the first and second half of the trajectories shows that there are no significant drifts along the experiments (Fig. S2). This implies that the traces are stationary and can reasonably be used for a comparison between temporal fluctuations in single cells and fluctuations across a population in a given time. The traces clearly show the instantaneous events of cell division and accumulation between them, which allows us to carry out below a detailed analysis of temporal trajectory features.

The distribution of fluorescence levels, representing the total amount of a specific protein in the cell, are extracted from several trajectories of individual trapped bacteria sampled every 3 minutes for about 70 generations, are shown in Fig. 2A, including three different proteins at 3 temperatures. It is seen that individual bacteria exhibit different protein distributions, and in particular their means are shifted with respect to each other. However, when plotted in scaled units, the distributions of individual cells extracted from their long-term temporal dynamics, collapse on top of one another (Fig. 2B). Moreover, they depict the same shape as the one measured for a snapshot of a large population (black line) described in [11]. In addition, the means and variances of all traces exhibit the same quadratic relationship previously observed for different populations (Fig. 2C). These results show that the universal statistical properties of protein variation reported in a preceding study [11] and measured from cell population snapshots, match the statistical properties of single-cell protein traces along time.

The individual protein traces, such as those depicted in Fig. 1, exhibit complex dynamics with characteristics on different time scales. Understanding the contribution of the different features to the universal statistical properties of protein variation is our main objective in what follows. Careful examination of the single-cell traces (Fig. 1C,D) reveals that they are dominated by the following features:

1) Each trace is composed of continuous portions (cell-cycles) separated by sharp drops at cell division (Fig. 3A, arrows). At each division event, the protein copy number is divided symmetrically between the two daughter cells (see Fig. S3). Within a cell-cycle, although small fluctuations exist (partly reflecting measurement noise; see Fig. S4), accumulation is smooth and can be well fitted by an exponential function, whose rate varies from one cell-cycle to the next (Fig. 3A and B).

2) The duration of the cell-cycle also varies between cycles.
3) The exponential accumulations of protein during each cell-cycle "ride" on top of a slowly-varying baseline (Fig. 3A, $a_k$s), representing the amount of protein in the cell at the beginning of the cell cycle following division.

These features change from one cell-cycle to the next (see Fig. S5) and can therefore contribute to the observed variation in protein content as well as the relationship between variance and mean. To disentangle the contributions of each of these features, we generate a simplified 3-parameter representation of the protein (and cell-size) traces:

$$p_k(t) = a_k \exp(\alpha_k t)$$

Where $p_k(t)$ is the amount of the specific protein under consideration in the cell during cycle $k$ at time $t$ measured from the preceding division. $a_k$ is the amount of protein in the cell at the beginning of cycle $k$ immediately after division, and $\alpha_k$ is the rate of exponential protein accumulation in the cell during cycle $k$. The time $t$ here ranges from 0 to $T_k$, where $T_k$ is the duration of cell-cycle $k$. This simplified representation of the dynamics accurately approximates our measurements, and preserves the important universal statistical properties of the protein variation discussed earlier (Fig. 3).

Using this simplified parametrization we can now quantitatively evaluate the contribution of each of the three parameters' variability to the universal statistical properties. To this end, we systematically reduce the effect of each of these features. Initially we remove the variability in the cell-cycle durations ($T_k$, see Fig. 4A) by setting them all to be constant and equal to the measured average. The resulting protein trace is therefore composed of a collection of exponentials with a baseline and exponential rates that vary between cycles yet all have the same cell-cycle duration. The protein distribution resulting from this manipulation is depicted by the black circles in Fig. 4B and is seen to be very similar to the distribution of the original measured protein trace. The conclusion from this procedure is that the variability in cell-cycle time contributes very little to the protein distribution shape.

Similarly in Fig. 4D, the baseline variable ($a_k$, see Fig. 4C) is substituted by its average, while keeping the other variables as measured. Here, the tail of the distribution is weakly affected but the lower end is modified in a manner easily understood: if all exponential functions start from exactly the same initial value, this is the minimal value in the sample. Moreover because of the increasing steepness of the exponential function, a uniform sampling in time results in a high sampling of the lower values of fluorescence, and the distribution will be strongly peaked at this value. This artificial lower cut-off on the fixed-baseline trace causes the deviation at the lower end of the curve. To further support this claim, we add to the constant $a_k$ a small Gaussian random value at each cycle with a standard deviation to mean ratio of 0.1 (smaller than the 0.34 measured), reproducing a distribution shape very similar to the original distribution (Fig. 4 red line). This indicates that the slow transgenerational variation in this parameter does not contribute to the shape of the distribution, and that the dissimilarity observed before when substituting a constant for the baseline is indeed an effect of the artificial bias induced by the substitution.

In contrast, substituting the exponential rates ($\alpha_k$, see Fig. 4E) by their average, while keeping the measured values of $T_k$ and $a_k$, changes the shape of the universal distribution (Fig. 4F). Neither the typical exponential-like tail nor the rounded lower-end is reconstructed by this model. This implies that the distribution of exponential rates among cell-cycles is crucial for shaping the protein distribution. This claim is further supported by the fact that keeping $T_k$ and $a_k$ constant, while replacing the measured exponential rates by random values, with similar distribution to the measured one, leaves the protein distribution intact (Fig. S6).

The analysis so far has treated each of the three variables separately; however Fig. 5(A – C) clearly shows that they are correlated with one another across cycles. To test for the contributions of these correlations, we construct from the measured set of variables a shuffled set, namely: each variable separately has the same distribution but they are not matched to one another correctly. The resulting distribution from the surrogate protein trace is shown in Fig. 5D by black circles. It is seen that, although the correlations between variables are relatively small, they affect the protein distribution shape (see discussion below). The nature of these correlations and their significance are subject to current and future investigation by our team.

**Discussion**

In studies of protein variability in cell populations, practically all experimental data were collected from single-cell measurements in large populations at a given time. In contrast, all models of protein variability and much of the interpretation attached to the measurements draw from a picture of protein dynamics along time in single cells: bursts in gene expression, cell growth and division along time, etc. While mRNA expression dynamics have been measured [30], the relevance of these measurements to our problem is limited due to the short timescales and the small correlation between mRNA and protein content [7]. Thus, until now no direct measurement of protein dynamics over long timescales have been analyzed to characterize their temporal statistical properties and to identify inheritance among multiple generations (although in [29] a sample trace of protein density was shown). Consequently the implicit question of the relationship between a cell population sample and a protein trace along time has remained largely open. In the current study we present such measurements for unprecedented extended timescales and address this question by direct comparison between these new temporal data on isolated bacteria and the corresponding population measurements.

The main result these data have revealed is that the universal statistical properties reported previously for populations of bacteria and yeast are also observed for the temporal dynamics of protein level in a single bacterium. Specifically, the shape of the protein distribution in scaled units has the same characteristic non-Gaussian shape measured in populations, and the variance shows a quadratic dependence on the mean. The match between individual and population distributions in scaled units shows that the single-cell explores, within <70 generations, the space of relative protein fluctuations with the same frequencies observed in a large population snapshot. The use of scaled fluctuations (common in Statistical Mechanics), had previously revealed that the protein distribution shape is universal in cell populations across two microorganisms and under a broad range of conditions. The results presented here demonstrate that this universal distribution is a reflection of the single cell dynamics at least in the case of bacteria. Currently, the lack of analogous temporal data prevents testing the generalization of this result to other cell types.

The second significant result is that the protein traces can be accurately described by only three parameters – the amount of protein in the cell at the beginning of the cell-cycle ($a_k$), the rate of protein accumulation in the cell ($\alpha_k$), and the cell-cycle time ($T_k$). Thus, the entire stochastic characteristics are accurately extracted from random variables drawn only once per cell cycle. This representation preserves the statistical properties, namely the distribution shape of cellular protein content and the relationship between its variance and mean. It shows that the relevant timescale for stochastic effects underlying protein distributions of highly expressed proteins is *the entire cell cycle*; on the short times between cell divisions they accumulate continuously in an almost deterministic manner, similar to the entire cell mass. This results in a timescale separation between fast, sub-generation processes such as transcription, translation, promoter states etc.; and the longer term trans-generational processes. Characterizing the entirety of intracellular processes within a cycle by a single rate parameter does not mean that these processes are deterministic. Rather, possibly due to their multiplicity, complexity and correlations, the minimal timescale over which significant changes appear is the entire cell cycle; faster processes are buffered from this level of organization.

This buffering is an important phenomenon that merits further study, both experimentally and theoretically, as it lies at the heart of understanding how one level of organization gives rise to the next level and determines its properties and functionality. In the context of protein variation the buffering of the fast sub-generation processes by the slower cellular organization can account for the insensitivity of the protein distributions to the intracellular processes. In addition, the observed universality of the distributions between populations of bacteria and yeast and across a wide range of biological realizations (different proteins and different metabolic conditions) suggests that similar buffering exists in yeast as well, and calls for further investigation of this phenomenon in other organisms and cell populations.

The compact parametrization of the protein trace by 3 variables per cycle enabled us to evaluate the contribution of each variable to the total protein variability along the trace. It was found that among the three parameters, two – $a_k$ and $T_k$ – can be substituted by constant values with minimal effect on the resulting protein distribution. On the other hand, the existence of a range of exponential rates is crucial for the generation of this distribution (Fig. 4B); their precise values are of lesser importance. Given, however, that in reality all three variables are random, correlations between them ensure that the effective range of exponentials is manifested in the trace and ensures the distribution shape .Thus it is important to understand the origin of the smooth exponential increase within a cell cycle, its variation from one cell cycle to the next, and the correlations among multiple phenotypes of the same cell.

If protein content were an isolated variable, its exponential accumulation during the cell-cycle would indicate that protein production rate is proportional to its current amount. However, our results show the same universal distribution and similar exponential accumulation for both proteins that strongly contribute to metabolism (LacO in lactose medium) and proteins that do not contribute (viral λ-phage promoter). The emerging conclusion is the entanglement of cellular processes underlying protein production leads to these dynamics, independently of the process details. This picture is consistent with the universal distribution found among populations in a broad range of biological realizations and protein functionality. We further note that recent theoretical work suggests that arbitrary complex networks of chemical interactions can give rise to effective exponential growth when projected on a single degree of freedom [31].

The entanglement of cellular processes suggests the existence of correlations between protein production and growth, regardless of the specific role of that protein in metabolism. Indeed, we find that exponentials can be fit also to the cell size dynamics of Fig. 1C (Fig. S7), consistent with previously published microscopy measurements showing that cell mass increases exponentially [16,32]. Fig. 6 shows that these exponents exhibit strong correlation with those of protein accumulation on a cycle-by-cycle basis for three different types of expression systems: regulated and metabolically relevant (LacO promoter), constitutive and metabolically irrelevant (ColE1-P1 promoter), and completely foreign to the bacteria (viral λ-phage promoter).

A testable prediction stemming from these results is that the rates of accumulation of different functionally non-related proteins within a cell cycle should be correlated with one another across cycles. Although this correlation is not expected to be uniform for all protein pairs, nevertheless it should extend beyond the level of specific regulatory modules to include correlations through the global metabolic network. Previous work has measured genome-wide pair correlations in yeast in snapshot measurements that sample the individual cells at different cell cycle stages [33]. The correlations are expected to be much stronger between the rates of production that represent the metabolic state of the cell throughout the entire cycle. These predicted pair correlations are expected to break down for low copy-number proteins which might then become sensitive to a specific intracellular process, such as transcription or translation [34].

Finally, our measurements show that the mean fluorescence for each trace, which reflects the average protein content in the cell over the measurement time, is different from cell to cell. Because of the long timescales associated with modulations of the average, and because our measurements are in arbitrary units, further experiments are required to calibrate its absolute value and to collect sufficiently large statistical samples to verify this individuality and clarify its source. Previous work has shown that slowly varying population averages exhibit nontrivial dynamics that can be highly significant biologically [26,35,36]. The question of these slowly-varying fluctuations, and in particular the relationship between temporal and population fluctuations in this non-universal regime, therefore remains a topic of high interest for future investigations.

## Methods

*Experimental setup and data acquisition*

Wild type MG1655 *E. coli* bacteria, expressing green fluorescent protein (GFP) from a medium copy-number plasmid (~15) under the control of the regulated Lac Operon (LacO) promoter, or the constitutive ColE1-P1 promoter, or the viral λ-phage promoter, were grown over night at 30°C, in M9 minimal medium supplemented with 1g/l casamino acids and 4g/l lactose (M9CL, for cells expressing GFP under the control of LacO), or 4g/l glucose (M9CG, for cells expressing GFP under the control of ColE1-P1 or λ promoter). The following day, the cells were diluted in the same medium and regrown to early exponential phase, Optical Density (OD) between 0.1 and 0.2. When the cells reached the desired OD, they were concentrated into fresh medium to an OD~0.3, and loaded into a microfluidic trapping device. The device consisted of multiple channels of 50 or 30 μm long, with 1 μm width and 1 μm height. The channels were closed at one end and open at the other, where large (30 x 30 μm) channels cross them perpendicularly (see Fig. 1). The cells were allowed to diffuse into the narrow channels and fresh M9CL (or M9CG for ColE1-P1 and λ promoter) medium was then flown through the large perpendicular channels to supply the thin perpendicular channels with nutrients to support the growth of trapped cells (Fig. 1A). The cells were allowed to grow in this device for ~70 – 100 generations, while maintaining the temperature at the required temperature, using a made-in-house incubator. A similar setup was developed independently by another group and used to follow cell size in other studies [29].

Images of the channels were acquired every 3 minutes in phase contrast and fluorescence modes using a Zeiss Axio Observer microscope with a 100x objective. This resolution ensures a continuous measurement relative to the typical timescales of change in both cell size and protein content, while minimizing the damage to the cells. The size and protein content of the mother cell (the cell at the closed end of the thin channel) were measured from these images using the image analysis software microbeTracker [37]. These data was then used to generate traces such as those presented in Figs. 1C and 1D, and for further analysis as detailed in the main text.

*Parametric representation of protein traces and the effect of the different parameters*

Traces of the total fluorescence in individual cells as a function of time were obtained from the acquired images. In each trace, the division points were identified and the time difference between two consecutive divisions was computed to find the cell-cycle time ($T_k$). The total fluorescence values between each two divisions were fit to an exponential function using two fitting parameters: the amount of protein at the beginning of each cycle ($a_k$), and the rate of exponential accumulation of protein ($\alpha_k$). These parameters were then used to reproduce the surrogate traces and calculate their statistical properties in order to compare to the statistical properties of the data. To assess the effect of each parameter on the statistical properties of each trace, new surrogate traces were produced in which the parameter(s) of interest (either $T_k$, $a_k$, or $\alpha_k$), which naturally varies between cycles, was replaced with a constant equal to its average along the trace. The other parameters were kept as obtained from the original fit, and the resulting new trace was used to calculate the new statistical properties and compare to those of the original data.

**Acknowledgements**

This work was supported by a US-Israel Binational Science foundation grant as well as by the Israel Science Foundation (Grant No. 1566/11, NB) and by the National Science Foundation (Grant PHY-1401576, HS). We are grateful to Omri Barak and Kinneret Keren for helpful discussions and critical reading of the manuscript.

## Figures

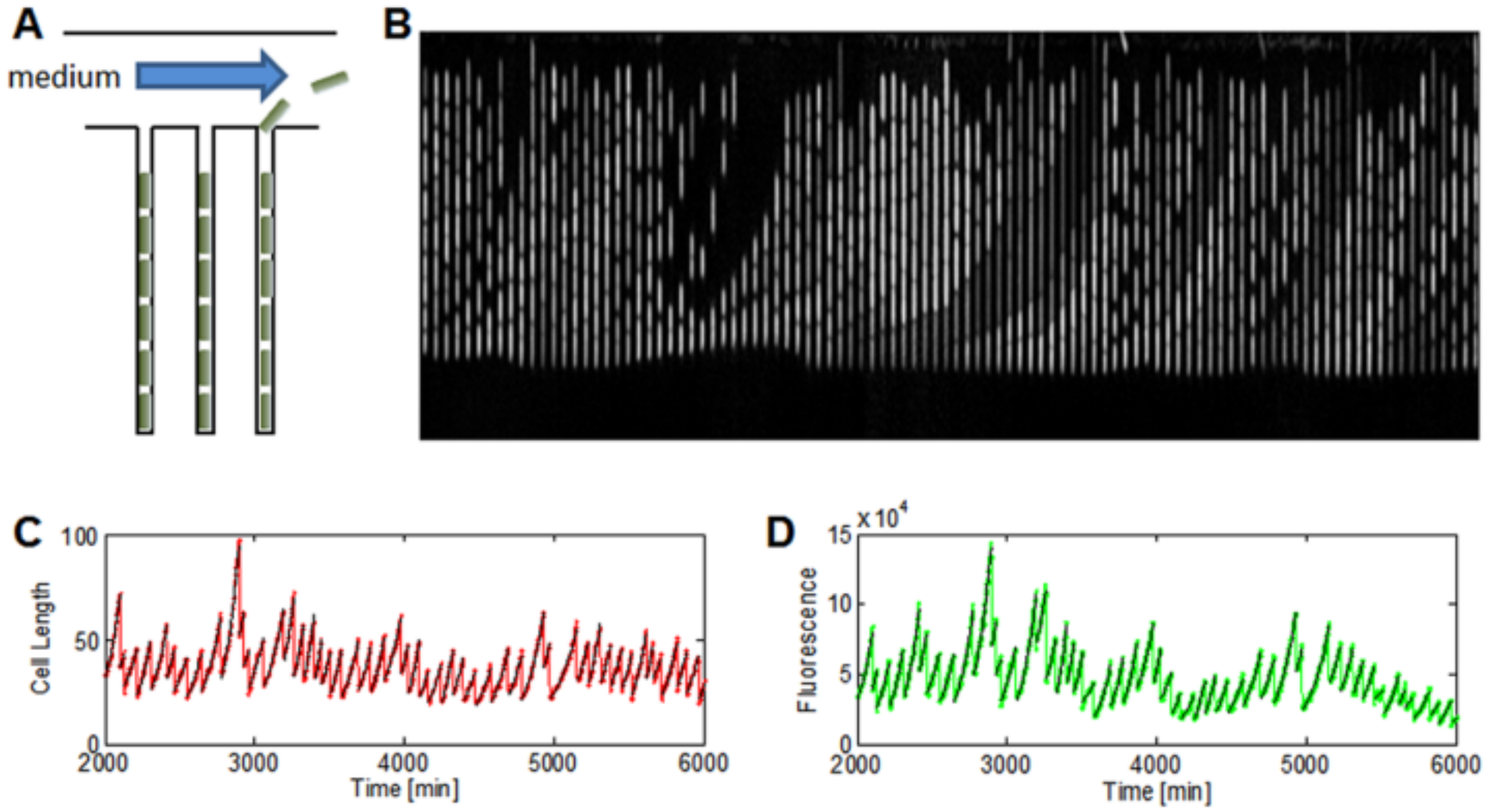

**Fig 1: Experimental setup and phenotypic traces of individual trapped bacteria.** (A) Schematics of the experimental setup: an array of channels (~ 30μm x 1μm x 1μm) closed at one end and open at the other, microfabricated in PDMS, designed for trapping individual bacteria. Fresh medium pumped through perpendicular channels feeds the trapped cells and washes out newly produced cells. Time-lapse images in phase contrast and fluorescence mode are acquired every 3 minutes using an inverted microscope. (B) A sequence of fluorescence images of a single channel with trapped bacteria at different times. The channel extends in the y-direction showing several bacteria. Subsequent time points (at 30 minute intervals) extend in the x-direction. (C, D): Temporal traces extracted for the trapped mother cell, from images such as (B), for cell size (C; red dots, measured in pixels) and fluorescence of a reporter protein regulated by the LAC operon promoter (D; green dots). An exponential $x_k(t) = a_k e^{\alpha_k t}, 0 < t < T_k$ is fitted to the k-th trajectory portion between consecutive divisions (black line).

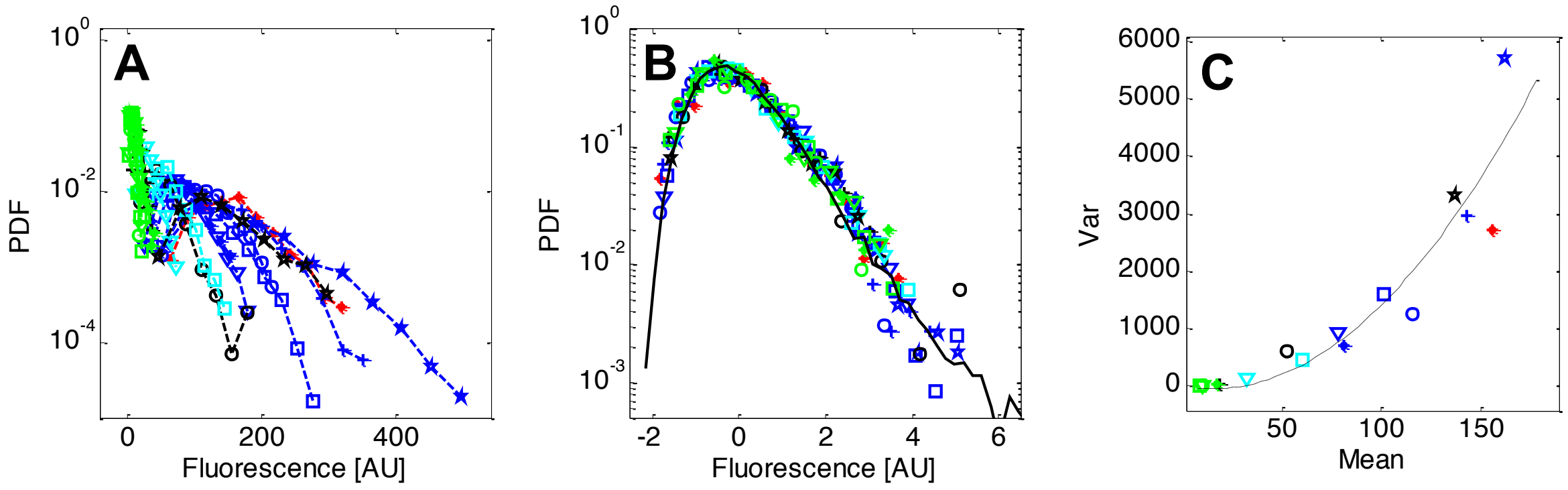

**Fig. 2. Universal features of fluctuations in temporal traces.** (A) Probability Density Functions (PDF) of fluorescence levels collected from traces such as Fig. 1D for 16 individual trapped cells, in which GFP is expressed from the highly induced LAC operon promoter (blue at 30°C, cyan at 28°C) or the constitutive ColE1-P1 promoter (red) or the λ-phage promoter (green). (B) Distributions of relative fluctuations for all the cells in (A): the x-axis is scaled by subtracting the mean and dividing by the standard deviation of each trace. For comparison, the scaled population snapshot distribution is shown by a black line (data from [11]; Lac operon promoter). (C) Variance as a function of mean for all measured trajectories. Colors and symbols are the same for all panels.

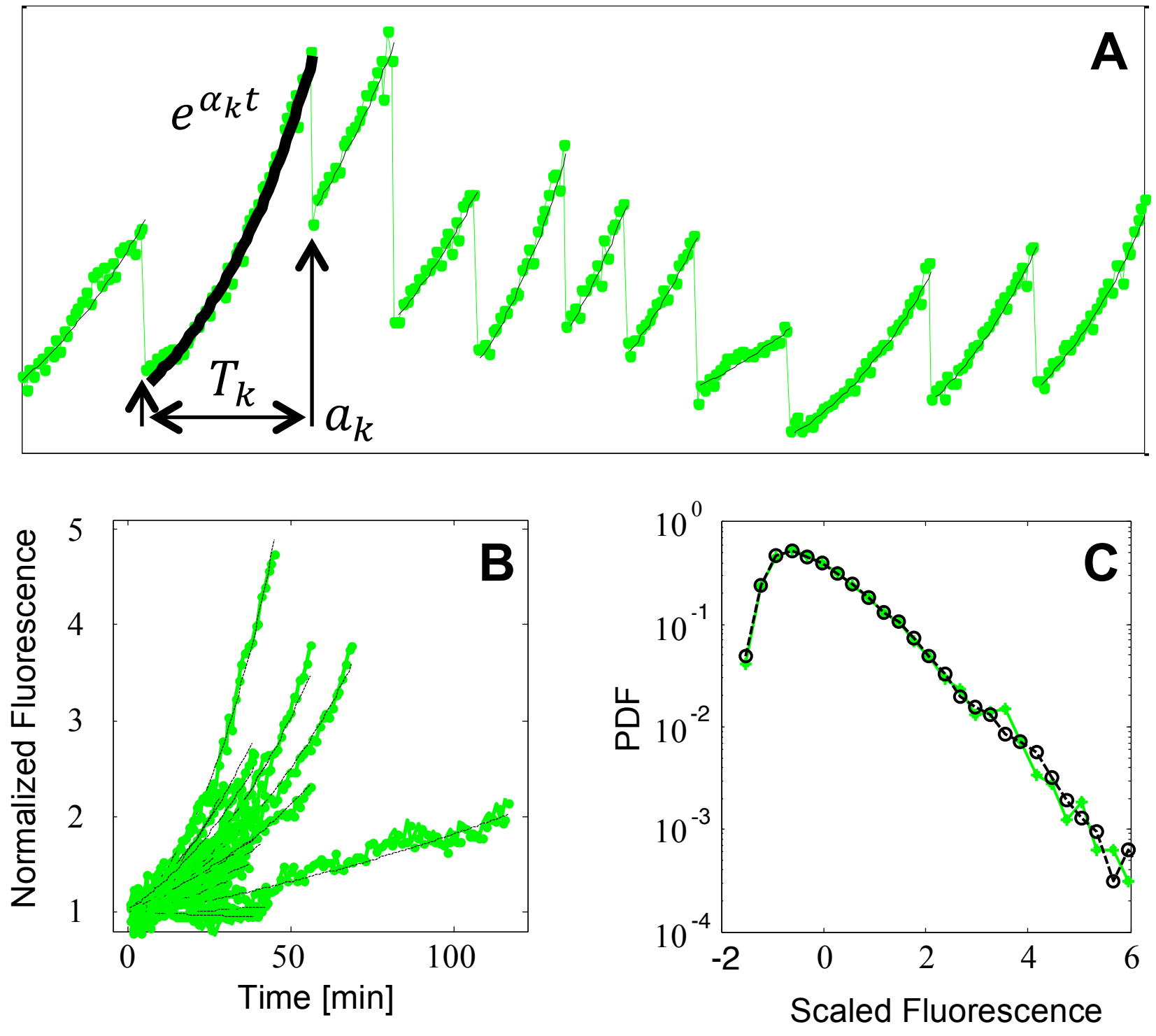

**Figure 3: Protein traces as a 3-variable random process.** (A) Protein traces are composed of discontinuous jumps (arrow; cell division events), exponential accumulation between divisions, and a slowly varying baseline (represented by $a_k$). (B) Continuous portions of the protein trajectories as a function of time between cell divisions (green). Time is aligned to the beginning of the cycle; protein level is normalized to be one at this initial time. Exponential functions $e^{\alpha_k t}$, $0 < t < T_k$, are fitted to the $k$-th cycle (black dashed lines). This plot highlights the significant variation in exponential rates $\alpha_k$ and in cycle times $T_k$ among cell cycles in one trace. Accounting also for the baseline $a_k$ in the fit results in the black dashed line drawn on the data points in (A). (C) The scaled distibution shape computed from the 3-variable process best fit to the data in each cycle, shown here by black circles, is indistinguishable from the raw data plotted in green.

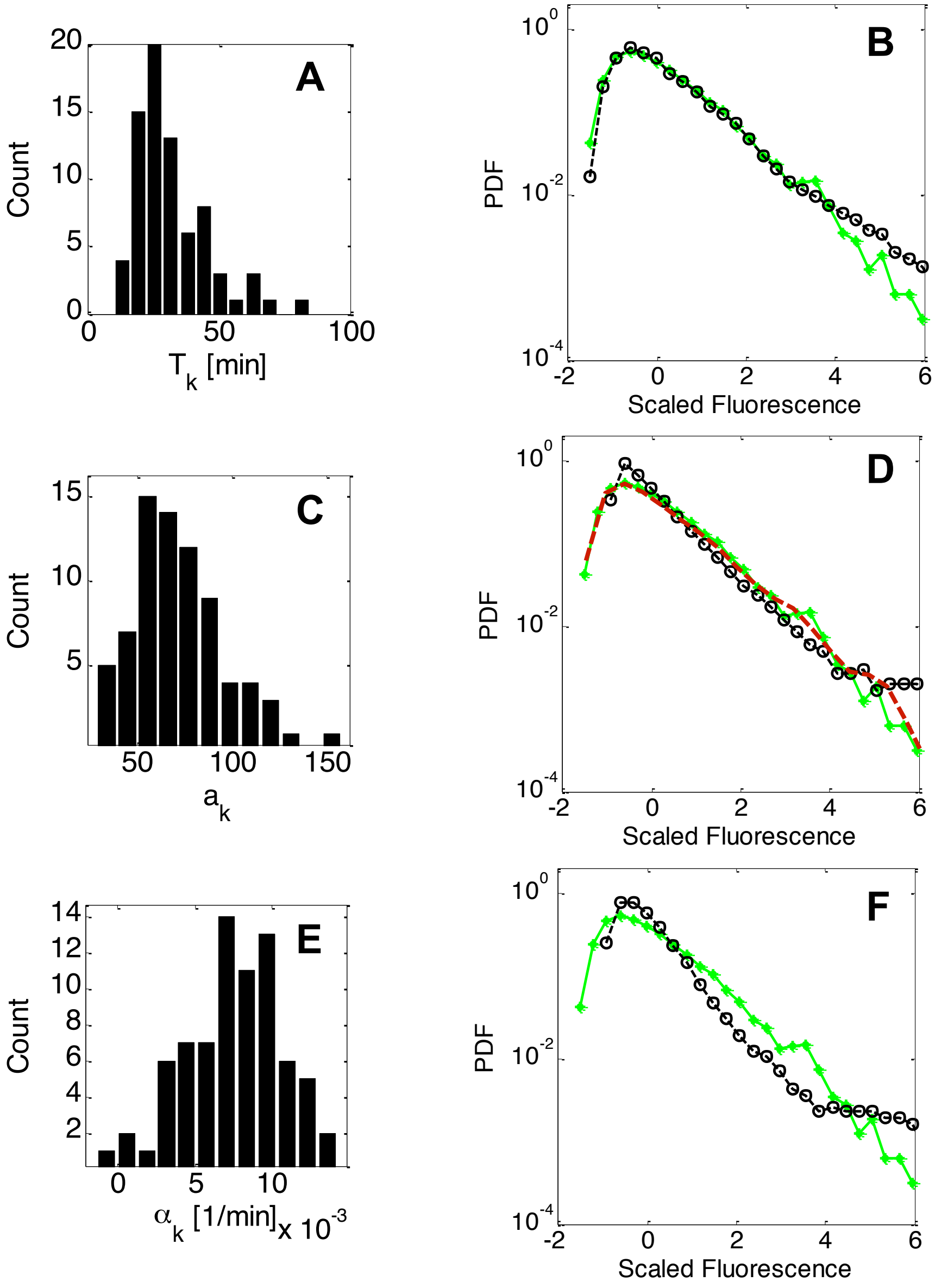

**Figure 4: Variation in the 3-variable approximation and its effect on the universal protein distribution**. The protein trace is approximated by a collection of *N* exponential functions of the form $\{a_k e^{\alpha_k t}, 0 < t < T_k\}_{k=1}^{N}$ .(A, C, and E) Histograms of the three parameters collected from different cell cycles in the same trace: (A) Times between cell divisions, $T_k$, with coefficient of vatiaion (CV) 0.42; (B) Baseline fluorescence level at the start of the cycle $a_k$, with CV 0.34 ; and (C) Exponential rates $\alpha_k$, with CV 0.42. In (B, D, and F) the measured distribution is compared to the distributions of surrogate traces, depicted by black circles, in which each of the random variables ((B) cycle durations, (D) baseline values, (F) exponential rates) was separately substituted by its average, and is thus constant along the trace, while the other two variables remain as measured. The red line in (D) depicts the distribution of a surrogate trace in which the baseline was substituted by a random value drawn from a Gaussian distribution around 1 with 0.1 standard deviation.

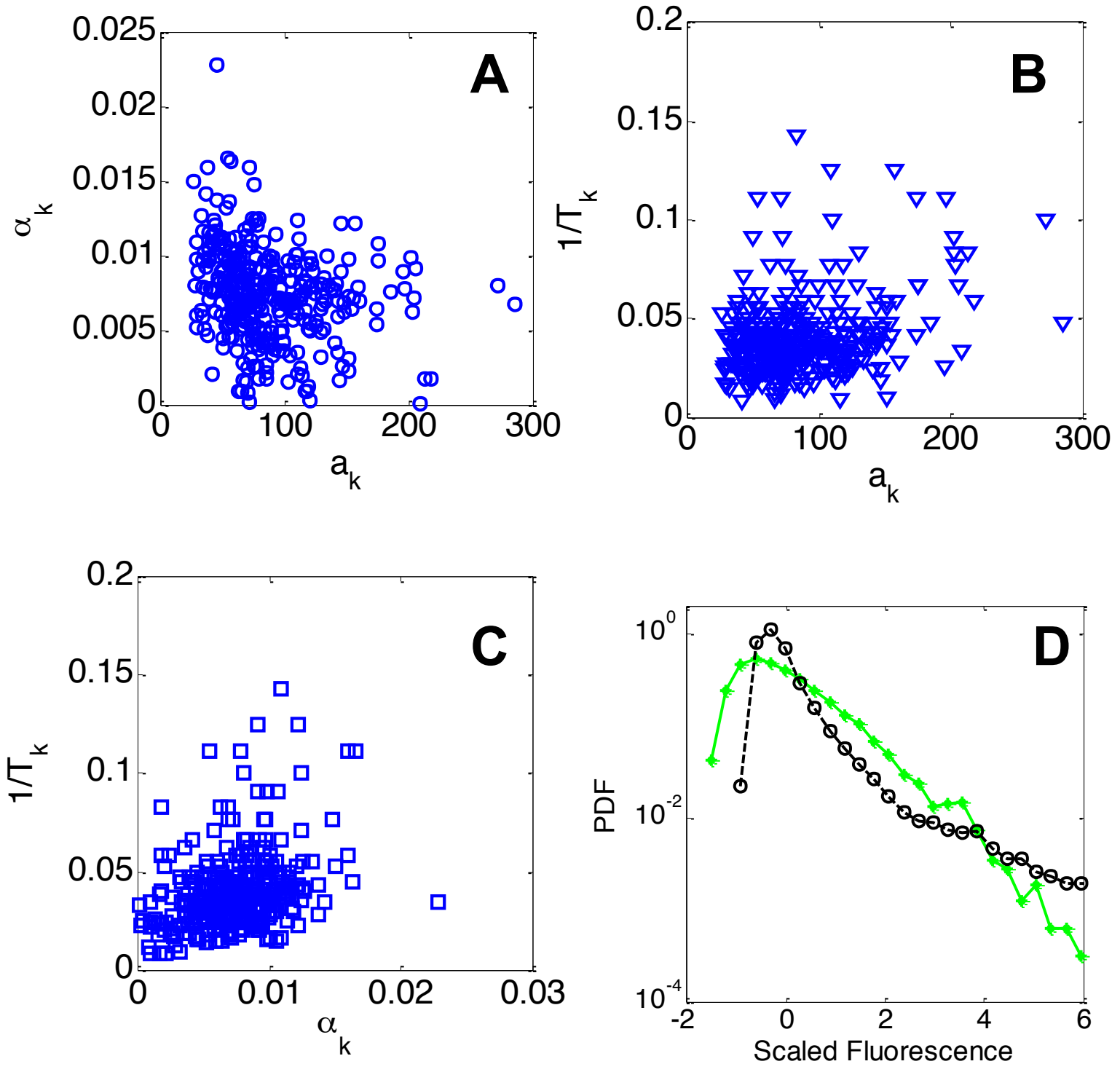

**Figure 5: Correlation in the 3-variable process and its contribution to universal distribution shape.** (A – C) Covariation of the three pairs of random variables across cycles in on individual trace. Correlation coefficients are -0.15, 0.49 and 0.29 respectively. (D) The collection of variables measured in one trace was shuffled such that their distributions are as measured but the correlations between their values at each cycle are destroyed.

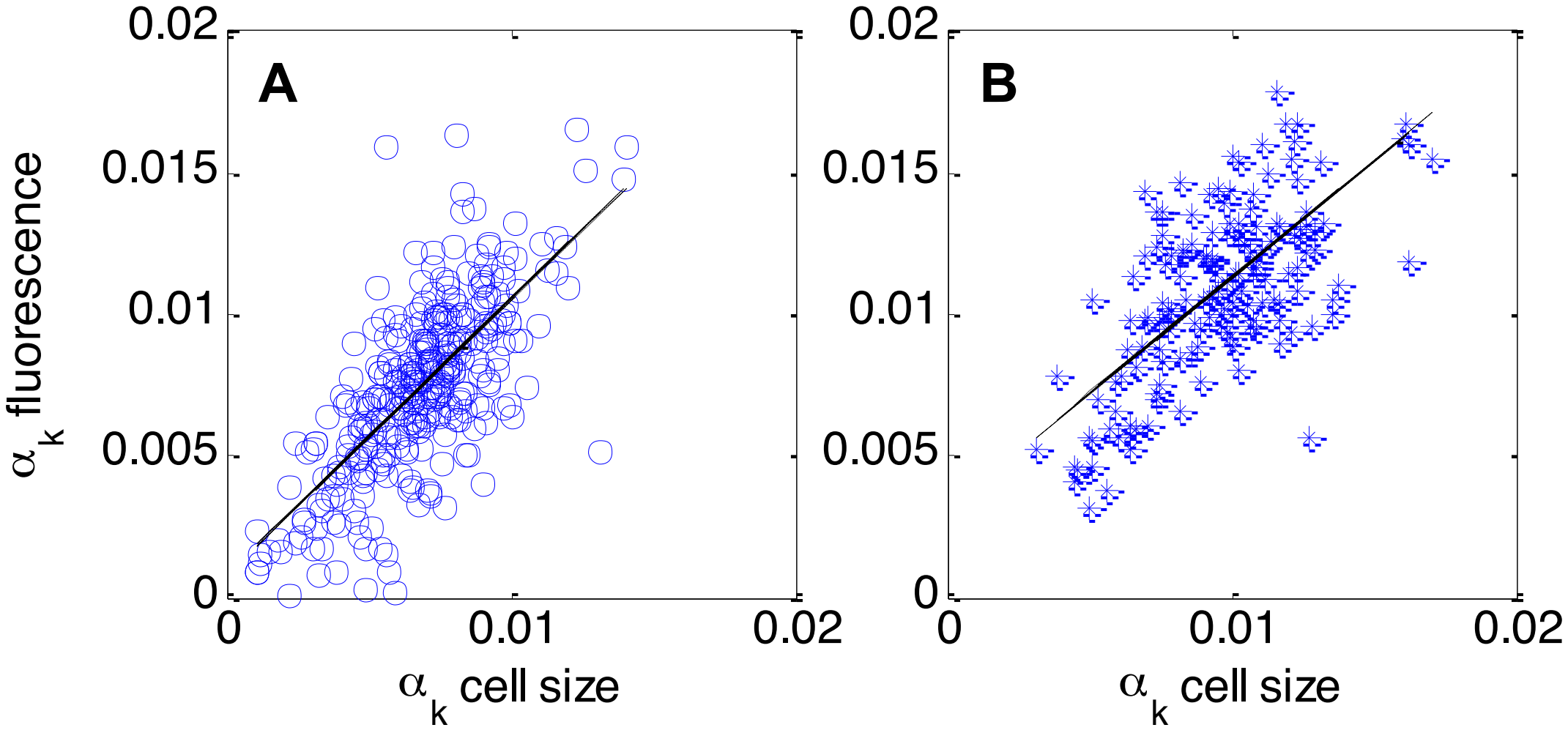

**Figure 6: Correlations between exponential rates of cell length and fluorescence.** For each cell cycle, the points represent the best fit exponential rate $\alpha_k$ to the cell length (x-axis) and to the fluorescence (y-axis). (A) Fluorescent protein is expressed from LacO promoter at 30°C, which is metabolically relevant in the lactose containing medium. (B) Fluorescent protein expressed from the λ-phage promoter which is entirely detached from cell metabolism. The correlation coefficients are 0.69 and 0.66, whereas the slopes of the best linear fits are 0.97 and 0.81, respectively.

## Supplementary Figures

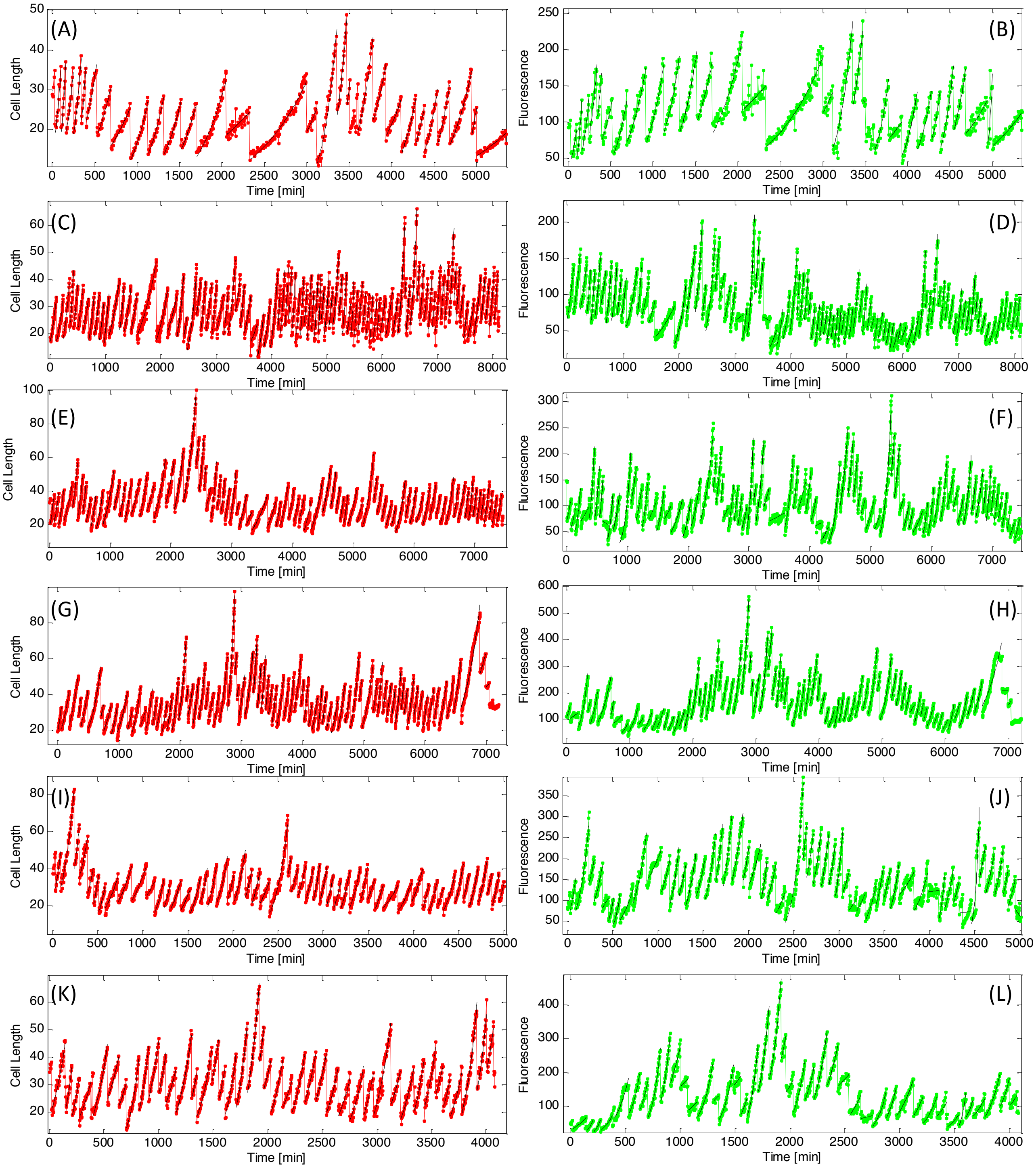

**Figure S1: Cellular phenotypic trajectories.** Trapped cells were followed over multiple generations, and their length (left) and fluorescence (right) reporting the gene E1P1 (A,B; measured every 6 min) and LacO (C-L; measured every 3 min) are plotted as a function of time. Black lines show separate exponential fits to the trajectory portions between cell divisions. Fig. 1C,D of the main text are samples extracted from panels G,H respectively.

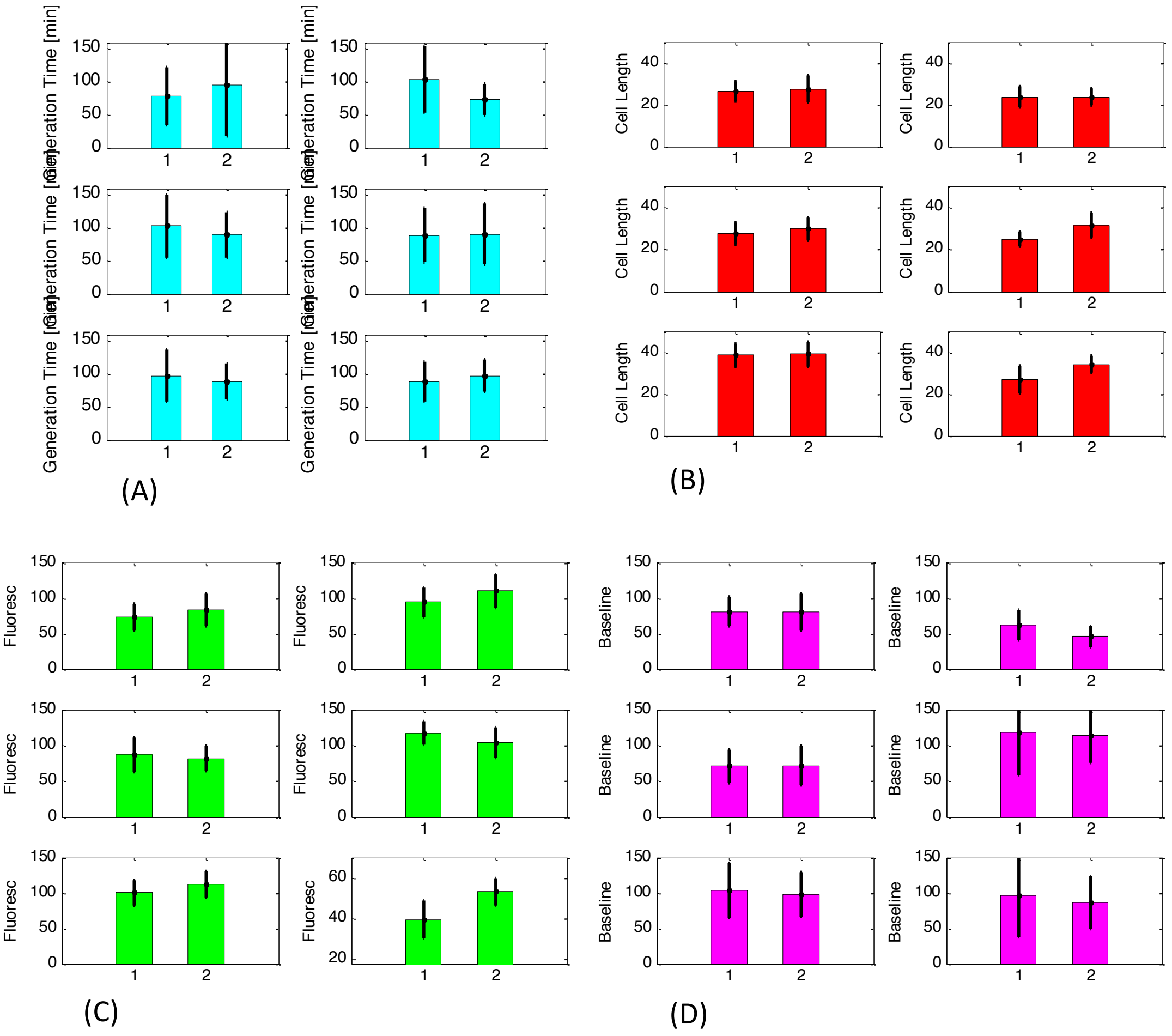

**Figure S2: Testing for stationarity of individual bacterial traces.** Time-averaged quantities for trapped cells were computed over the first and second half of each trajectory: (A) generation time; (B) cell length; (C) fluorescence; (D) baseline fluorescence, e.g. fluorescence value at the start of each cell cycle.

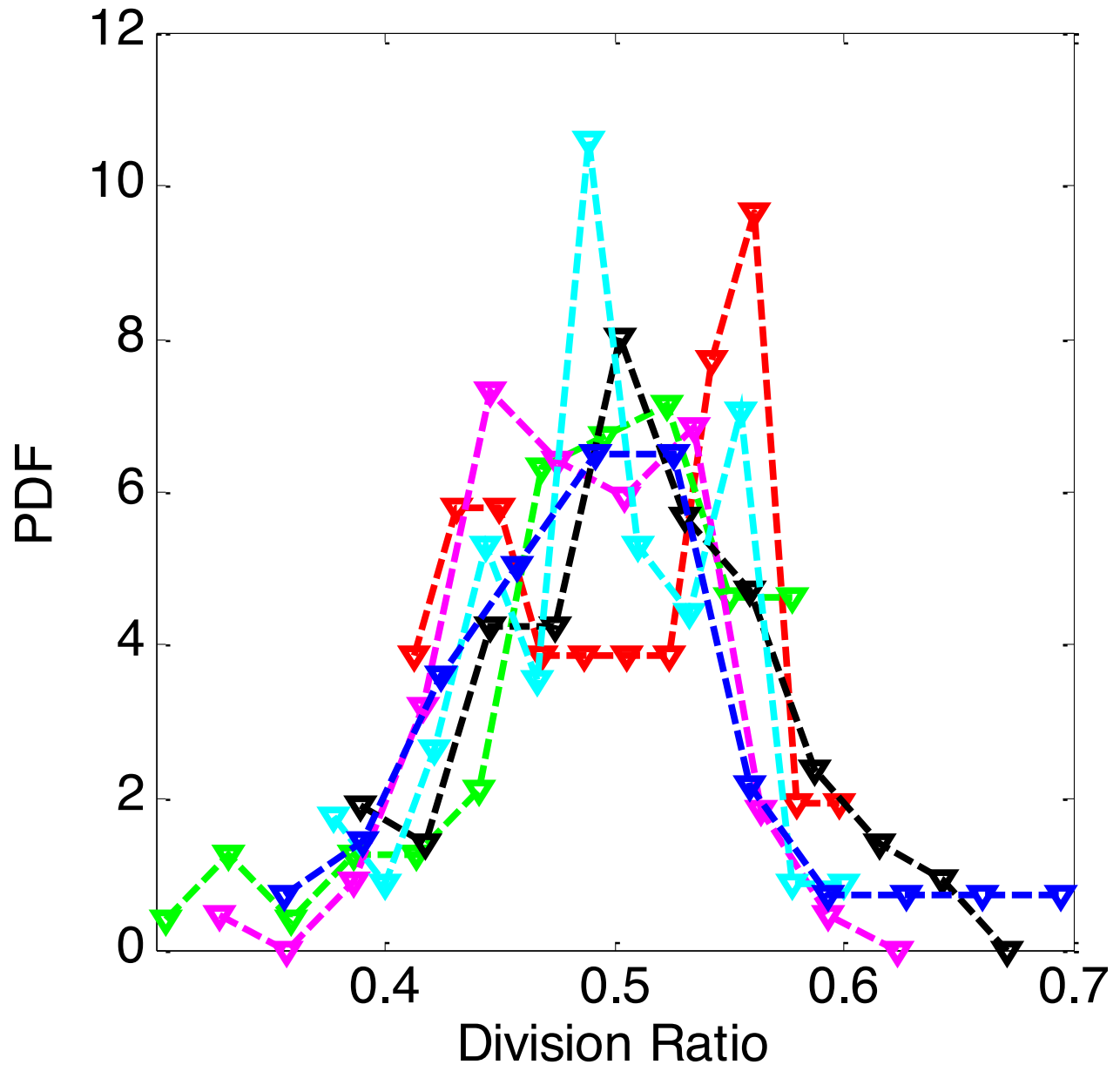

**Figure S3: Division of protein copy number between the two daughter cells.** The fluorescence intensity in individual bacteria was measured immediately before and after division and the ratio (after/before) was computed. The distribution of these division ratios over multiple cell divisions is plotted for different cells (colors). The results show that the distribution of division ratios is approximately Gaussian, with average 0.499 and standard deviation 0.063.

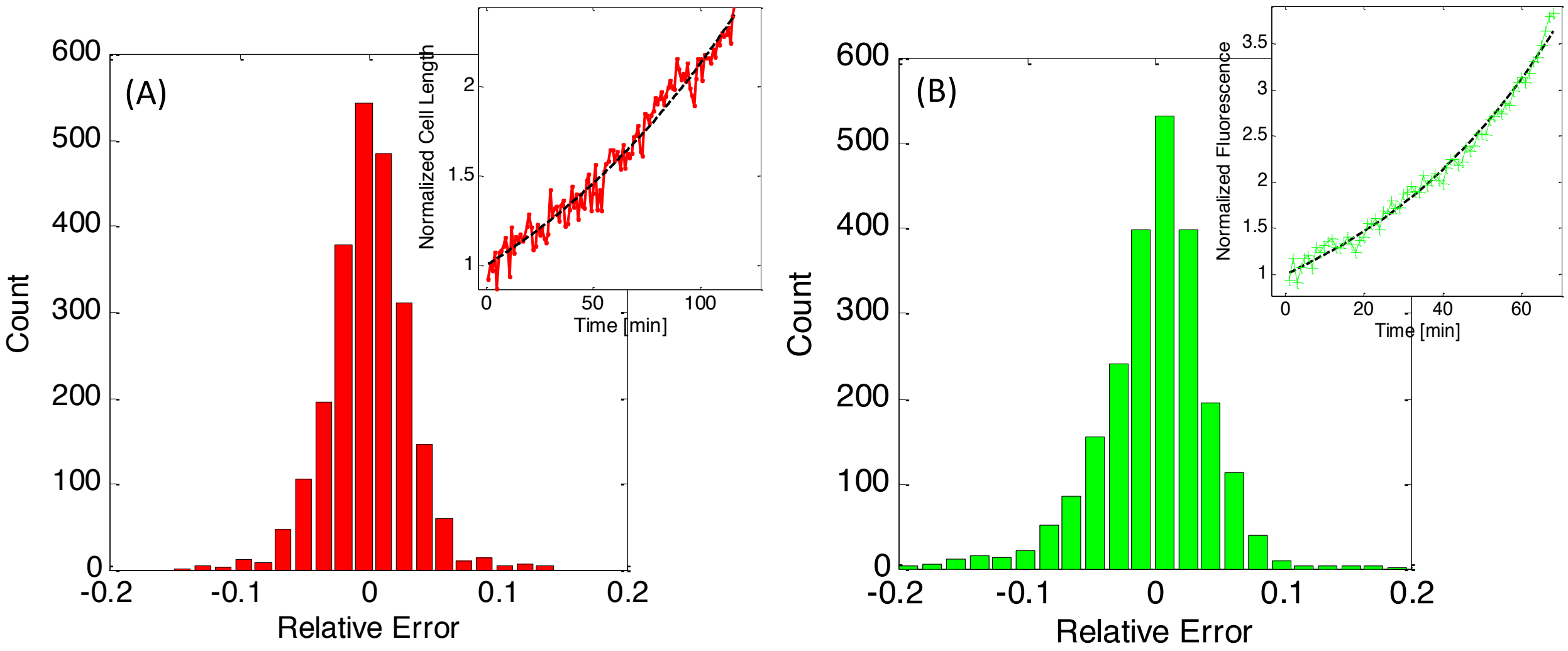

**Figure S4: Residual errors – difference between data and exponential fits.** After fitting each cycle by an exponential function, the difference between the measured data and the exponential fit was computed for cell length (A) and fluorescence (B) for one trajectory. These differences were normalized by the average of the entire trajectory and shown here is the histogram of these relative error values. Such distributions are typical to other trajectories as well. Insets show examples of single cycles, zooming on the actual data points as compared to the exponential fits. Same cell analyzed in Fig. 3 of the main text.

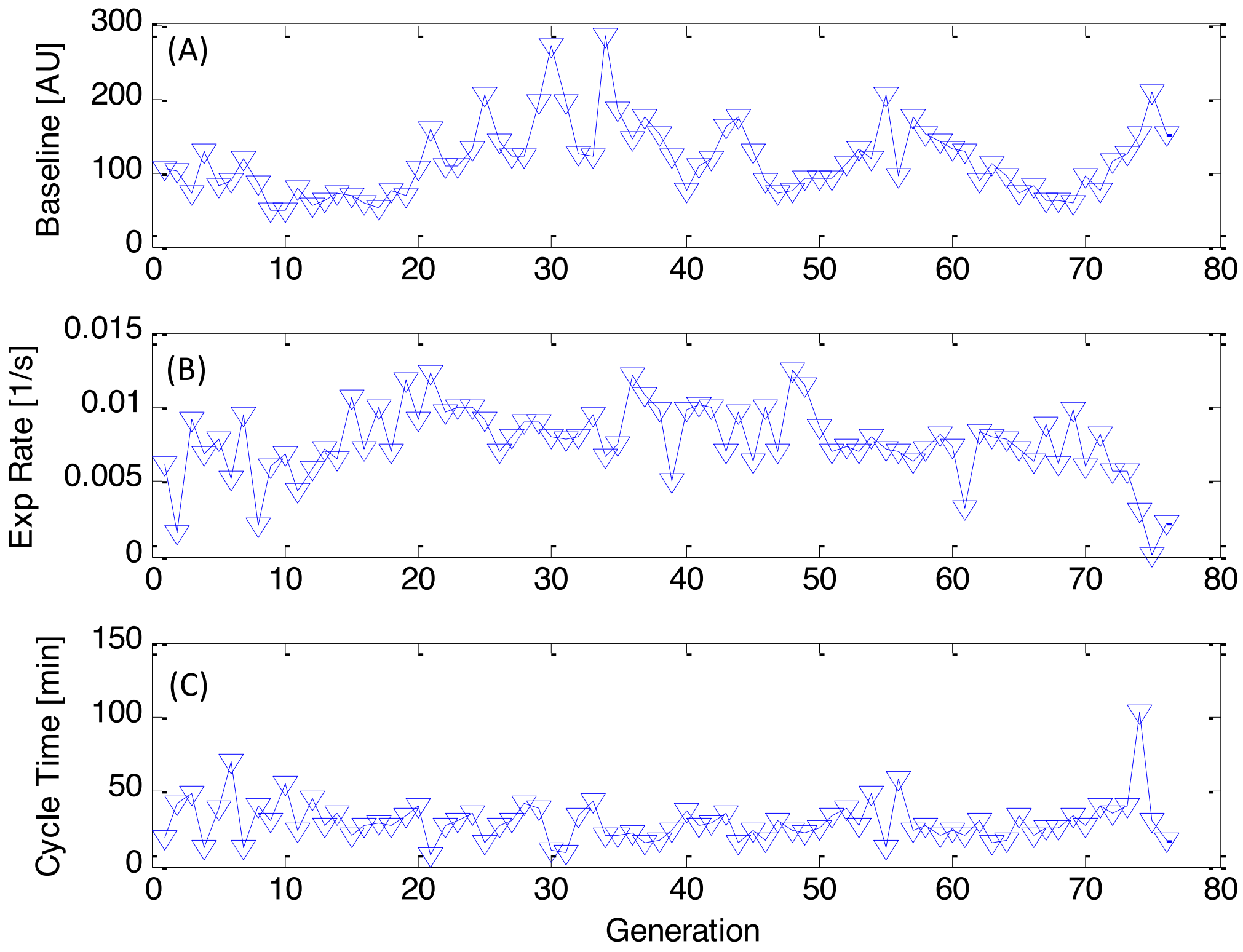

**Figure S5: Protein trace parameters along generations in one trapped cell.** (A) Baseline value of fluorescence at beginning of each cycle as a function of generation number. (B) Best fit exponential rate for fluorescence for each cycle as a function of generation number. (C) Cycle time duration in seconds as a function of generation number. Data are shown for the cell of Fig. 3 in the main text.

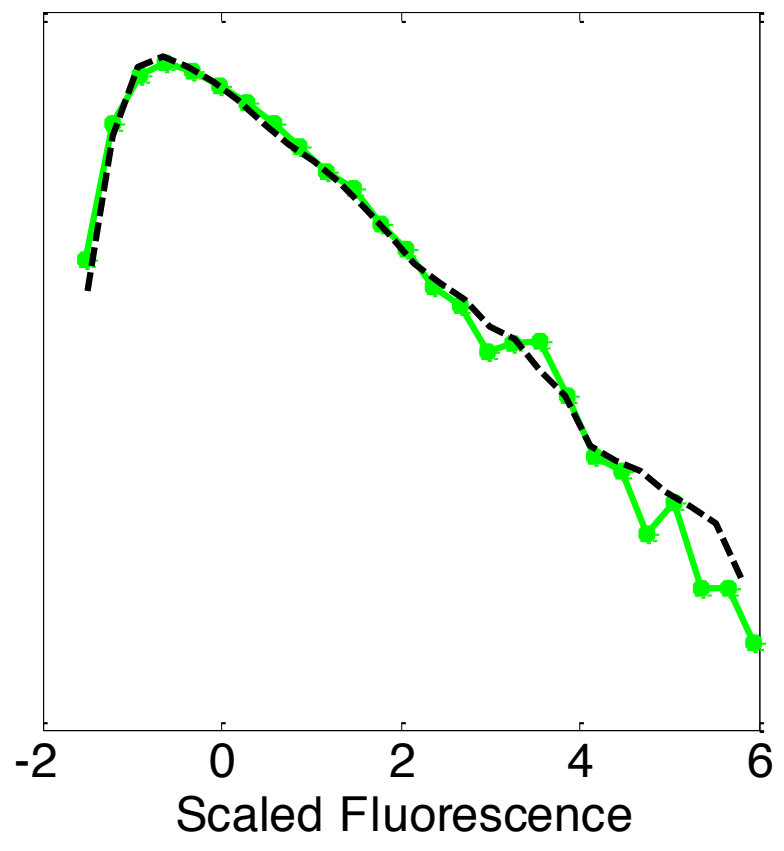

**Fig. S6. Random exponential rates shape the universal distribution**. The measured fluorescence distribution is shown in green. In black the distribution of the 3-parameter approximation, in which the baseline values were substituted by a random value drawn from a Gaussian distribution around 1 with 0.1 standard deviation, the cell cycle times substituted by a fixed duration, equal to the measured average (90 min), and the exponential rates substituted by random values drawn from a Gaussian with the measured mean and standard deviation (0.0235 $min^{-1}$ and 0.0085 $min^{-1}$ respectively) . The reduced model histogram was computed from the same number of points as in the experiment (~300 cycles).

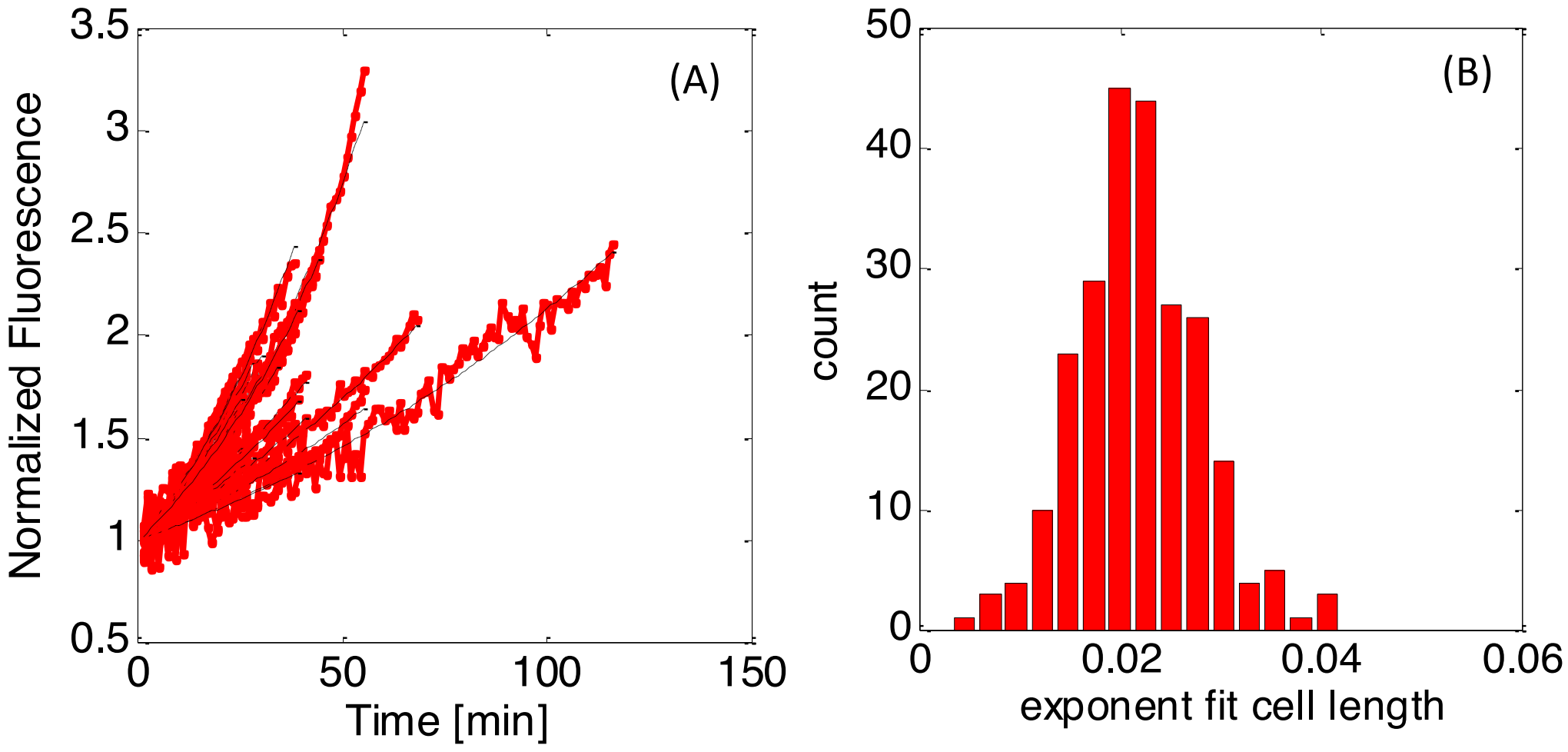

**Figure S7: Exponential fits to cell length trajectory portions between cell divisions.** (A) Several segments of cell length trajectories as a function of time between cell divisions, from Fig. S1A. Time is aligned to the beginning of the cycle; cell length is normalized to be 1 at this initial time. Exponential functions $e^{\alpha t}$ are fitted to the data (black dashed lines). (B) Histogram of the exponential rates α of cell length growth along the cellular trajectory. Data are shown for the cell of Fig. 3 in the main text.